\documentclass{article}
\def\beq{\begin{equation}}
\def\eeq{\end{equation}}
\def\bea{\begin{eqnarray}}
\def\eea{\end{eqnarray}}
\def\bq{\begin{quote}}
\def\eq{\end{quote}}

\def\nnb{\nonumber}
\def\ga{\left(}
\def\dr{\right)}
\def\aga{\left\{}
\def\adr{\right\}}

\def\rar{\rightarrow}
\def\lrar{\Longrightarrow}
\def\nnb{\nonumber}
\def\la{\langle}
\def\ra{\rangle}
\def\nin{\noindent}
\def\ba{\begin{array}}
\def\ea{\end{array}}
\def\bm{\overline{m}}

\def\b{\bullet}

\begin{document}
\topmargin -1.5cm
\oddsidemargin +0.2cm
\evensidemargin -1.0cm
\pagestyle{empty}
\begin{flushright}
PM/99-24\\
(revised July 1999) 
\end{flushright}
\vspace*{5mm}
\begin{center}
\section*{Strange-quark mass 
from combined $e^+e^-$  and $\tau$-decay data:
\\ test of the isospin symmetry and
implications on $\epsilon'/\epsilon$ and $m_{u,d}$}
\vspace*{1.5cm}
{\bf Stephan Narison}
\\
\vspace{0.3cm}
Laboratoire de Physique Math\'ematique et Th\'{e}orique\\
UM2, place Eug\`ene Bataillon\\
34095 Montpellier Cedex 05, France\\
E-mail:
narison@lpm.univ-montp2.fr\\
\vspace*{2.5cm}
{\bf Abstract} \\ \end{center}
\vspace*{2mm}
\noindent
I extract the strange-quark mass using a $\tau$-like decay
sum rule for the $\phi$-meson, and some other sum rules involving its difference with
the vector component of the hadronic $\tau$-decay. As a conservative estimate, one obtains 
to order $\alpha_s^3$: $\overline{m}_s$(1 GeV) = $(178\pm 33)$ MeV~$\lrar
~\overline{m}_s$(2 GeV) = $(129\pm 24 )$ MeV, while the positivity of the spectral function
leads to the upper bound: $\overline{m}_s(1 ~{\rm GeV})\leq (200\pm 28)~\mbox{MeV}~\Longrightarrow~
\overline{m}_s(2~{\rm GeV})\leq (145\pm 20)~\mbox{MeV}$. These results are in good
agreement with the existing sum rule and $\tau$-decay results, and, in particular,
with the result from the the sum
rule involving the difference of the isoscalar and isovector components of the $e^+e^-\rar$ hadrons data. This
signals small effects of the $SU(2)$ isospin violation due to the $\omega$--$\rho$ mixing parameters, and
 questions the reliability of the existing sum rule estimates of these parameters.  Combining our
result  with the recent data on
$\epsilon'/\epsilon$, we can estimate, within the standard model, the four-quark weak matrix elements
$B_6^{1/2}-0.54 B^{3/2}_8$ to be about $ (2.8\pm 1.3)$. This result may suggest a large violation of the vacuum
saturation estimate similarly to the case of the four-quark condensates obtained from
the sum rules analysis, and can serve as a guide for a future accurate
non-perturbative extraction of such matrix elements. Combining our result with the sum rule estimate of
$m_u+m_d$ and with the Dashen formula for the mass ratio, one can deduce the update values:
$\overline{m}_d(2~{\rm GeV})= (6.4\pm 1.1)~{\rm MeV}$ and 
$\overline{m}_u(2~{\rm GeV})= (2.3\pm 0.4)~{\rm MeV}$.

\noindent
\vspace*{5cm}
\begin{flushleft}
PM/99-24 \\
May 1999 \\
(revised July 1999)\\
\end{flushleft}
\vfill\eject
\pagestyle{plain}
\setcounter{page}{1}
\section{Introduction}
Among the most important parameters of the standard model and of chiral symmetry
breaking are the light quark masses. In addition to a much better understanding of the realizations
of chiral symmetry breaking \cite{LEUT0}, the recent measurement of the $CP$ violating parameters
$\epsilon'/\epsilon$
\cite{EPS} needs a much better control of the Standard Model predictions, which can be largely affected by the
value of the running strange quark mass within the widely used parametrization of the non-perturbative
four-quark matrix elements $Q_6$ and $Q_8$ \cite{BURAS}. A lot of effort reflected in the literature
\cite{PDG} has been put into extracting directly from the data the running quark masses
using the SVZ QCD spectral sum rules (QSSR) \cite{SVZ,SNB}, lattice data \cite{LATT}
and LEP experiments \footnote{For recent comparisons of different determinations, see e.g. 
\cite{LEUT} and \cite{SNR}.}. In this note, I propose some new $\tau$-like decay sum rules for the extraction
of the strange-quark mass from the $\phi$-meson, and $\tau$-decay data, which, unlike the sum rule proposed in
\cite{SN1} (hereafter referred as SN) involving the difference of the isoscalar and isovector component of the
$e^+e^-\rar$ hadrons data, are  not directly affected by the less (theoretically) controlled effect of the
$\omega$--$\rho$ mixing. We shall see later on that the result of the present analysis surprisingly agrees quite
well with the previous result in SN assuming $SU(2)$ isospin symmetry (neglect of the $\omega$--$\rho$
mixing effect). This fact then signals small effects of $SU(2)$ violation due to the 
$\omega$--$\rho$ mixing parameters in the sum rules involving the difference of the isoscalar and isovector
components of the $e^+e^-\rar$ hadrons data. It also questions the reliability of the existing sum rule
estimates of these parameters \cite{MALTA} following the pioneer work of SVZ \cite{SVZ2}, and the uses of
such estimates for extracting $m_s$ \cite{MALT}.
\section{Normalizations and notation}
We shall be concerned with the transverse two-point correlator:
\beq
\Pi^{\mu\nu}_{ab}(q)\equiv i\int d^4 x~ e^{iqx}\la 0|{\cal T} J^\mu_a(x)\ga
J^\nu_b(0)\dr^{\dagger}|0\ra=-(g_{\mu\nu}q^2-q_\mu q_\nu)\Pi_{ab}(q^2)~,
\eeq
built from the $SU(3)$ component of the local electromagnetic current:
\beq
J^\mu_{EM}=V^\mu_3(x)+\frac{1}{\sqrt{3}}V^\mu_8(x)~,
\eeq
where:
\beq
V^\mu_a(x)\equiv \sqrt{\frac{1}{2}}\bar{\psi}(x)\lambda_a\gamma^\mu {\psi}(x)~; 
\eeq
$\lambda_a$ are the diagonal flavour $SU(3)$ matrices:
\beq
\lambda_3=\sqrt{\frac{1}{2}}\ga
\ba{ccc}
1&&\\
&-1&\\
&&0\\
\ea
\dr ~,\,\,\,\,\,\,
\lambda_8=\sqrt{\frac{1}{6}}\ga
\ba{ccc}
1&&\\
&1&\\
&&-2\\
\ea
\dr~,
\eeq
acting on the basis defined by the up, down and strange quarks:
\beq
\psi(x)=\ga
\ba{c}
u(x)\\
d(x)\\
s(x)\\
\ea
\dr~.
\eeq
In terms of the diagonal quark correlator:
\beq
\Pi^{\mu\nu}_{jj}(q)\equiv i\int d^4 x e^{iqx}\la 0|{\cal T} J^\mu_j(x)\ga
J^\nu_j(0)\dr^{\dagger}|0\ra=-(g_{\mu\nu}q^2-q_\mu q_\nu)\Pi_{jj}(q^2)~\, \,
j=u,d,s~,
\eeq
the previous $SU(3)$ flavour components of the electromagnetic correlator 
read\footnote{We shall follow the normalization
used in SN \cite{SN1}.}:
\bea
\Pi_{33}&=&\frac{1}{2}.\frac{1}{2}\ga \Pi_{uu}+\Pi_{dd}\dr~,\nnb\\
\Pi_{88}&=&\frac{1}{2}.\frac{1}{6}\ga \Pi_{uu}+\Pi_{dd}+2\Pi_{ss}\dr~,\nnb\\
\Pi_{38}&=&\frac{1}{4\sqrt{3}}\ga \Pi_{uu}-\Pi_{dd}\dr~,\nnb\\
\eea
Therefore, the $e^+e^-\rar$ hadrons total cross-section reads:
\beq
\sigma(e^+e^-\rar{\rm hadrons})_{u,d,s}=\frac{4\pi^2\alpha}{s}e^2\frac{1}{\pi}
\aga \rm{Im} \Pi_{33}(s)+\frac{1}{3}\rm{Im} \Pi_{88}(s)+\frac{2}{\sqrt{3}}\rm{Im}
\Pi_{38}(s)\adr~.
\eeq
In a narrow-width approximation (NWA), the resonance $H$ contributions to the
spectral functions can be introduced through:
\beq
\la 0|V^\mu_a|H\ra=\epsilon^\mu\frac{M_H^2}{2\gamma_{Ha}}~,
\eeq
where the coupling $\gamma_{Ha}$ is related to the meson leptonic width as:
\beq
\Gamma_{H\rar e^+e^-}=\frac{2}{3}\alpha^2\pi\frac{M_H}{2\gamma_{Ha}^2}~,
\eeq
which is itself related to the total cross-section:
\beq
\sigma(e^+e^-\rar H)=12\pi^2\frac{\Gamma_{H\rar e^+e^-}}{M_H}\delta (s-M^2_H)~.
\eeq
\section{$\phi$-meson $\tau$-like decay sum rules}
We shall be concerned with the hypothetical $\tau$-like decay sum rules :
\beq
R_{\tau,\phi}\equiv\frac{3|V_{ud}|^2}{2\pi\alpha^2}S_{EW}\int_0^{M^2_\tau}
ds\ga 1-\frac{s}{M^2_\tau}\dr^2\ga 1+\frac{2s}{M^2_\tau}\dr\frac{s}{M^2_\tau}
\sigma_{e^+e^-\rar \phi,\phi',...}~,
\eeq
and the $SU(3)$-breaking combinations:
\beq
\Delta_{1\phi}\equiv R_{\tau,1}-R_{\tau,\phi},
\eeq
and \cite{SN1}: 
\beq
\Delta_{10}\equiv R_{\tau,1}-3R_{\tau,0}~,
\eeq
which involves the difference of the isoscalar ($R_{\tau,0}$) and isovector ($R_{\tau,1}$) sum
rules \`a la Das-mathur-Okubo \cite{DMO},
and vanishes in the $SU(3)$ symmetry limit.
Here, \cite{MARC} $S_{EW}=1.0194$ is the electroweak correction, $|V_{ud}|^2=0.975$
is the CKM mixing angle \footnote{One should notice that this term is an overall factor appearing in both
sides of the sum rules. It has been only introduced in order to mimic the expression of the true
$\tau$-decay width in the charged current sector.}.
\\ Its QCD expression is known to be
\cite{BNP} (hereafter referred as BNP):
\beq
R_{\tau,\phi}=\frac{3}{2}|V_{ud}|^2 S_{EW}\frac{2}{9}\ga
1+\delta_{EW}+\delta^{(0)}+\sum_{D=2,4,6}{\frac{\delta^{(D)}_{ss}}{M^D_\tau}}\dr~,
\eeq
where \cite{LI}:
\beq
\delta_{EW}=0.0010~.
\eeq
Using fixed-order perturbation theory (FOPT), one obtains partly from BNP:
\bea
\delta^{(0)}&=&\ga \frac{\alpha_s}{\pi}\equiv a_s\dr
+5.2023a_s^2+26.366 a_s^3\pm 134 a_s^4~,\nnb\\
\delta^{(2)}_{ss}&=&-2.1a_s\lambda^2-12\overline{m}_s^2\ga
1+\frac{13}{3}a_s+30.4a_s^2+(d_3\equiv k_3-45.147)a_s^3
\pm d_4 a_s^4-2\frac{a_s\lambda^2}{M^2_\tau}\dr~,\nnb\\
\delta^{(4)}_{ss}&=&a_s^2\Bigg{[}\frac{11}{4}\pi\la \alpha_s G^2\ra-36\pi^2\la
m_s\bar ss\ra-8\pi^2\sum_{q= u,d,s} \la m_q\bar qq\ra\Bigg{]}+36
\overline{m}^4_s~,\nnb\\
\delta^{(6)}_{ss}&\simeq& 7\frac{256}{27}\rho \alpha_s\la \bar ss\ra^2~.
\eea
The expressions of the $m_s^2$ terms have been derived to order $a_s^2$ in SN and \cite{MALT}.
The order $a_s^3$ terms have been obtained by \cite{CHET0}, using the results in \cite{CHET1}. The other
higher dimension terms come from SN using the BNP's result.
We have added, into the expression, the contribution of the tachyonic gluon mass $\lambda^2$ 
motivated to mimic the unknown large order terms (renormalons,...) of the QCD perturbative asymptotic series
\cite{VALYA},
where \cite{CNZ,SN2}:
\beq
a_s \lambda^2\simeq -(0.06\sim 0.07)\rm{GeV}^2~.
\eeq 
In order to be conservative, we shall consider this effect as a source of systematic errors.
One can estimate \cite{CHET0} the constant term $k_3a_s^3$ of the two-point
correlator, by assuming a ``na\"\i ve" geometric growth of the perturbative coefficients, as usually done
(BNP, \cite{KATA,SNT}). Assuming a 50\% error in this
estimate, one then obtains:
$k_3 \approx 24.1415^2/2.6667\simeq (218.554\pm 109.268)$. This leads to the value:
\beq
d_3=173.4\pm 109.2~,
\eeq
of the full coefficient of the $a_s^3$-term.
 Alternatively, a much more na\"\i ve estimate of the $a_s^3$ coefficient $d_3$ can be obtained from an
assumed geometric growth of the perturbative coefficients of the $R_{\tau,\phi}$ perturbative series, after
the uses of the contour integral. In this way, one obtains, a value $d_3\simeq 30.410^2/4.333\simeq 208$,
which is consistent with the previous number. Therefore, we consider the error estimate given
previously, as a generous estimate. We add to this error estimate, the one due to the unknown $a_s^4$,
which we have again estimated from a na\"\i ve geometric growth of the perturbative coefficients
(134 for $\delta^{(0)}$ and $d_4= d_3^2/30\approx 1002$ for $\delta^{(2)}_{ss}$).\\ The
expression of the running coupling to three-loop accuracy can be parametrized as
\cite{SNB}:
\bea
a_s(\nu)&=&a_s^{(0)}\Bigg\{ 1-a_s^{(0)}\frac{\beta_2}{\beta_1}\log
\log{\frac{\nu^2}{\Lambda^2}}\nnb \\
&+&\ga a_s^{(0)}\dr^2\Big{[}\frac{\beta_2^2}{\beta_1^2}\log^2\
\log{\frac{\nu^2}{\Lambda^2}}-\frac{\beta_2^2}{\beta_1^2}\log
\log{\frac{\nu^2}{\Lambda^2}}-\frac{\beta_2^2}{\beta_1^2}
+\frac{\beta_3}{\beta_1}\Big{]}+{\cal{O}}(a_s^3)\Bigg\}~,
\eea
with:
\beq
a_s^{(0)}\equiv \frac{1}{-\beta_1\log\ga\nu/\Lambda\dr}~,
\eeq
and 
$\beta_i$ are the  ${\cal{O}}(a_s^i)$ coefficients of the 
$\beta$ function in the $\overline{MS}$-scheme for $n_f$ flavours, 
which read, for
 three flavours:
\beq
\beta_1=-9/2~,~~~~~\beta_2=-8~,~~~~~\beta_3=-20.1198~.
\eeq
$\Lambda$ is a renormalization-group-invariant (RGI)
 scale, but is 
renormalization-scheme-dependent.
The expression of the running quark mass in terms of the
invariant mass $\hat{m}_i$ is \cite{SNB}:
\bea
\bm_i(\nu)&=&\hat{m}_i\ga -\beta_1 a_s(\nu)\dr^{-\gamma_1/\beta_1}
\Bigg\{1+\frac{\beta_2}{\beta_1}\ga \frac{\gamma_1}{\beta_1}-
 \frac{\gamma_2}{\beta_2}\dr a_s(\nu)~\nnb \\
&+&\frac{1}{2}\Bigg{[}\frac{\beta_2^2}{\beta_1^2}\ga \frac{\gamma_1}
{\beta_1}-
 \frac{\gamma_2}{\beta_2}\dr^2-
\frac{\beta_2^2}{\beta_1^2}\ga \frac{\gamma_1}{\beta_1}-
 \frac{\gamma_2}{\beta_2}\dr+
\frac{\beta_3}{\beta_1}\ga \frac{\gamma_1}{\beta_1}-
 \frac{\gamma_3}{\beta_3}\dr\Bigg{]} a^2_s(\nu)~\nnb \\
&+&1.95168a_s^3\Bigg\}~,
\eea
where the $a_s^3$ term comes from \cite{VELT}; $\gamma_i$ are the ${\cal{O}}(a_s^i)$ coefficients of the 
quark-mass anomalous dimension, which read
for three flavours:
\beq
\gamma_1=2~,~~~~\gamma_2=91/12~,~~~~\gamma_3=24.8404~.
\eeq
In the present analysis, we shall use as inputs \cite{PDG}, \cite{BETHKE}--\cite{OPAL}:
\beq
\Lambda=(375\pm 50)~\rm{MeV}~,
\eeq
and \cite{SNx,SNB}:
\bea
\la \alpha_s G^2\ra&\simeq& (0.07\pm 0.01)~\rm {GeV}^4~,\nnb\\
m_{u,d}\la \bar uu\ra&\simeq& - \frac{1}{2} m^2_\pi f^2_\pi~,\nnb\\
\la \bar ss\ra/\la \bar uu\ra&\simeq& (0.7\pm 0.1)~,\nnb\\
\rho \alpha_s\la \bar uu\ra^2&\simeq& (5.8\pm 0.9 )\times 10^{-4}~\rm{GeV}^6~.
\eea
For the phenomenological estimate of the $R_{\tau,\phi}$ sum rule, 
we introduce, as stated before, the contributions of the resonances using a NWA,
where we use \cite{PDG}:
\beq
\Gamma_{\phi(1019)\rar e^+e^-}= (1.32\pm 0.04)~ {\rm keV}~,\,\,\,\,\,
\Gamma_{\phi(1680)\rar e^+e^-}= (0.48\pm 0.14)~ {\rm keV}~.
\eeq
\begin{table*}[hbt]
\begin{center}
\setlength{\tabcolsep}{1.2pc}
\caption{Phenomenological estimates of $R_{\tau,I}$, $\Delta_{1\phi}$ and central
values of $\overline{m}_s$ to order $\alpha_s^3$}
\begin{tabular}{c c c c c c}
\hline 
 &&&&&\\
$M_\tau$ &$\frac{9}{2}R_{\tau,\phi}$&
$\overline{m}_s$(1 GeV)&$R_{\tau,1} $&$
\Delta_{1\phi}$&$\overline{m}_s$(1 GeV)\\
~[GeV]&&[MeV]&from
\cite{ALEPH2}&&[MeV]\\ &&&&&\\
\hline
&&&&&\\
1.2& $0.85\pm 0.04$&223.3&$2.03\pm 0.025$&$1.18\pm0.05$&222.5\\
1.4&$1.47\pm 0.06$&186.1&$1.95\pm 0.025$&$0.48\pm0.07$&193.1\\
1.6&$1.59\pm 0.05$&173.1&$1.85\pm 0.025$&$0.26\pm 0.06$&185.8\\
1.8&$1.54\pm 0.07$&207.6&$1.78\pm 0.025$&$0.24\pm 0.07$&226.5\\
&&&&&\\
\hline 
\end{tabular}
\end{center}
\end{table*}
\begin{table*}[hbt]
\begin{center}
\setlength{\tabcolsep}{1.2pc}
\caption{Different sources of errors in units of [MeV] for the estimate of $\overline{m}_s$ (1 GeV)}
\begin{tabular}{c c c}
\hline 
&& \\
Sources&$|\Delta {m_s}|$ from $R_{\tau,\phi}$&$|\Delta {m_s}|$ from $\Delta_{1\phi}$\\
&&\\
\hline
&&\\
\underline{Experiment}&&\\
&&\\
Data&19&26\\
Mixing angle $(35\pm 1)^0$&0.8&0.8\\
&&\\
Total exp. (quadratic)&19&25\\
&&\\
&&\\
\underline{Theory}&&\\
&&\\
$\Lambda=(375\pm 50)$ MeV&15&8\\
$\pm 134\alpha_s^4$&11&--\\
$\pm 12m_s^2\times 109.3\alpha_s^3$&6&6\\
$\pm 12m_s^2\times 1002\alpha_s^4$&6&8\\
FOPT or CIE series&8&13\\
tachyonic gluon mass&15&2\\
$\la \alpha_s G^2\ra=(0.07\pm 0.01)$ GeV$^4$&0.04&0\\
$\la \bar ss\ra/\la \bar uu\ra = 0.7\pm 0.1$&6&4\\
$d=6$ condensates &3&2\\
&&\\
Total theory (quadratic)&27&19\\
&\\
\hline 
\end{tabular}
\end{center}
\end{table*}
\\
The continuum contribution (which is relatively small) has been estimated using a SU(3) symmetry
relation on the results obtained in SN for the isoscalar sum rule. The sum of the different
contributions is given in Table 1. \\
In our estimate, we include the $m_s^2\alpha_s^3$ term correction with the coefficient estimated previously,
which is in the range of the value obtained for the charged current sector \cite{PICH,ALEPH}. However, we
expect that the estimate of the error done in this way can be an overestimate of the real theoretical
error.\\ Another eventual source of error is the small deviation from the ideal mixing angle
$\theta_V=35.3^0$, which experimentally is
\cite{PDG}:
$\theta_V\simeq 35^0$ \footnote{Some theoretical estimates from the off-diagonal mass matrix leads to
slightly higher values of the mixing angle \cite{PDG}.}. We take into account this deviation by assuming
that it introduces an error of about
$0.01\times R_{\tau,\phi}$ on the value of $R_{\tau,\phi}$ given in Table 1. This effect is much smaller
than the experimental error in the estimate of $R_{\tau,\phi}$ and $\Delta_{1\phi}$ given in Table 1, and
leads to a small error (see Table 2).
\section{$\overline{m}_s$ from $R_{\tau,\phi}$: estimate and upper bound}
 Using the previous inputs, we give to order $\alpha_s^3$, the
central value of
$\overline{m}_s$(1 GeV) from $R_{\tau,\phi}$ in the 3rd column of Table 1. We consider, as an
optimal result, the one obtained for $M_\tau=1.6$ GeV, where the estimate presents a minimum, which
is an indication of a compromise region where both the continuum + higher-states contributions and
the non-perturbative + higher-order terms of the QCD series to the sum rules
are (reasonably) small. The different sources of errors for this value are given in Table 2.
One may check the convergence of the QCD perturbative series at this scale $M_\tau$=1.6 GeV. Using the
corresponding value of $a_s(1.6$ GeV) $\simeq 0.114$, one can deduce each terms of the perturbative
series :
\bea
1+\delta^{(0)}&\simeq& 1+0.114+0.068+0.04 \pm 0.02+...\nnb\\
\delta^{(2)}_{ss}|_{m_s}&\simeq&-12\overline{m}_s^2\Big{[} 1+0.494+0.390+(0.283\pm 0.178) \pm
0.169+...\Big{]}~,
\eea
which shows a slow (but not a catastrophic) convergence. The last error is due to the
estimate based on the geometrical growth of the coefficients of the unknown $a_s^4$ term
in the series. The error $\pm 0.178$ is the assumed uncertainty of the $d_3$
coefficient. However, the convergence of the QCD series is much better here than in
the case of the charged currents
\cite{ALEPH,PICH}. In the present analysis, we consider that the remaining terms of the
series can be mimiced by the tachyonic gluon mass introduced previously in \cite{VALYA,CNZ},
where a detailed explanation of the physical role of this non-standard term is discussed.
In order, to avoid a misinterpretation of our result, we include again this effect into
the theoretical error, where a such tachyonic mass contribution is mainly important in the
unflavoured term and tends to increase the mass of the strange quark by about 9\%, while
it is almost negligible of about 1\% in the flavoured $SU(3)$ breaking terms
\cite{CNZ}. Within the previous four different sources (including the one due to the treatment
of the QCD series: fixed order (FOPT) \cite{BNP} or contour improved expansion (CIE)
expected to improve the convergence of the QCD series \cite{LEDI}) of theoretical
uncertainties which we have included (see Table 2), we expect that the theoretical
errors given in Table 2 are largely overestimated. Taking into account our previous
discussions and the fact that the signs of each
$a_s$ corrections are all positive, the truncated PT series is expected to be a quite good
approximation of the full series. One should also remember that, for the quantity of
interest $m_s$, the corrections are about half of the one appearing in the PT series above.  From the
previous analysis, and after solving a quartic equation in $m_s$, the
$R_{\tau,\phi}$ sum rule leads to the result (to order $\alpha_s^3$) given in Table 1 in
the entry corresponding to $M_\tau=1.6$ GeV, where we only give the
central value of the strange quark mass coming from a mean value of the one from FOPT and CIE
(here, the result from CIE
is slightly higher than from FOPT by 10 MeV). Collecting all different sources of  errors in
Table 2, we consider, as a final estimate:
\beq\label{rfi}
\overline{m}_s(1 ~{\rm GeV})\simeq (173\pm 19_{exp}\pm 27_{th})~\mbox{MeV}~.
\eeq
We can also use the positivity of the spectral function and retains only the lowest ground state
$\phi$-meson contribution in order to derive an upper bound for $m_s$. The absolute (weakest) bound comes from
the inclusion of the tachyonic gluon contribution into the QCD expression and using the CIE. It reads:
\beq\label{bound}
\overline{m}_s(1 ~{\rm GeV})\leq (200\pm 17_{exp}\pm 22_{th})~\mbox{MeV}~,\,\,\,\,\,\Longrightarrow
\,\,\,\,\,\overline{m}_s(2~{\rm GeV})\leq (145\pm 12_{exp}\pm 17_{th})~\mbox{MeV}~,
\eeq
which is, however, stronger than the one obtained in \cite{YND} from the pseudoscalar moment sum rules.
\section{$\overline{m}_s$ from $\Delta_{1,\phi}$}
We use the more recent and precise value of $R_{\tau,1}$ from Fig. 17 of ALEPH \cite{ALEPH2} (see also OPAL
\cite{OPAL}),
which agrees within the errors with the one estimated in SN and in \cite{PICH2,SNT} from $e^+e^-\rar$
hadrons  data. Then, we consider the $SU(3)$-breaking $\tau$-like sum rule:
\beq
\Delta_{1\phi}\equiv R_{\tau,1}-R_{\tau,\phi},
\eeq
which, as the one in SN: 
\beq
\Delta_{10}\equiv R_{\tau,1}-3R_{\tau,0}~,
\eeq
which involves the difference of the isoscalar and isovector spectral function \`a
la Das-mathur-Okubo \cite{DMO},
vanishes in the $SU(3)$ symmetry limit. Its advantage 
over the previous $R_{\tau,\phi}$ sum rule is the vanishing of the $\lambda^2$ tachyonic gluon-mass 
contribution
to leading order, and its weaker sensitivity to the value of $\alpha_s$. Here, the
tachyonic gluon mass tends to decrease the strange quark-mass value by
1\% \cite{CNZ}.The advantage of this sum rule compared with the difference of sum rules proposed in SN,
is its weaker sensitivity to the strength of the
$SU(2)$ isospin violation due to the $\omega$--$\rho$ mixing contribution. The
phenomenological value of
$\Delta_{1\phi}$ is given in the 5th column of Table 1, while the central value of
$\overline{m}_s$ (1 GeV) resulting from an average between the FOPT and CIE results (here, the CIE result is
lower by about 26 MeV than the FOPT one) is
given in the last column of this table. As in the case of the
$R_{\tau,\phi}$ sum rule, the optimal result is obtained for $M_\tau \simeq 1.6$ GeV. The different
sources of errors in the estimate is given in Table 2.  The final result is:
\beq\label{best}
\overline{m}_s(1~{\rm GeV})\simeq (186\pm 25_{exp}\pm 19_{th})~\mbox{MeV}~.
\eeq
\section{$\overline{m}_s$ from $\Delta_{1,0}$: updated}
The two results from the two sum rules are in a good agreement with each others.  
This result also agrees with the previous estimate in SN from the difference $\Delta_{10}$ between the
isovector and isoscalar $\tau-$like decay sum rule in $e^+e^-$, to order $a_s^2$ \cite{SN1}:
\beq
\overline{m}_s(1~{\rm GeV})\simeq (194\pm 25_{exp}\pm 19_{th})~\mbox{MeV}~.
\eeq
obtained by assuming a good realization of the $SU(2)$ isospin symmetry, i.e. by neglecting
the
$\omega$--$\rho$ mixing violating term. We update the previous estimate of SN 
by including the $a^3_s$ contribution and by using the recent value of $R_{\tau,1}$ 
from ALEPH \cite{ALEPH2}
into Table 2 of SN. The strengths of the theoretical errors are about the same as the ones for $\Delta_{1,\phi}$
given in Table 2. In
this way, one obtains, to order
$a^3_s$:
\beq\label{d10}
\overline{m}_s(1~{\rm GeV})\simeq (176\pm 24_{exp}\pm 19_{th})~\mbox{MeV}~.
\eeq
The agreement of this result with the two former ones in Eqs (\ref{rfi}) and (\ref{best}) rules out definitely
unusual huge
$SU(2)$ violation induced by the $\omega$-$\rho$ mixing term advocated in \cite{MALT}, which can decrease
the previous value of $m_s$ by about a factor 2. We shall comment on this point in the next
sections.
\section{Final value and comparison with other results}
We take as our final result the mean value of the three different results given in Eqs. (\ref{rfi}),
(\ref{best}) and (\ref{d10}), and take as a conservative error, the largest one from the three
determinations, which is in Eq. (\ref{rfi}). In this way, we obtain:
\beq\label{final}
\overline{m}_s(1~{\rm GeV})\simeq (178\pm 33)~\mbox{MeV}~,\,\,\,\,\,\Longrightarrow
\,\,\,\,\,\overline{m}_s(2~{\rm GeV})\simeq (129\pm 24)~\mbox{MeV}~.
\eeq
This result agrees within the errors with the one from the
published results for the $\Delta S=-1$ component of
$\tau$-decay \cite{ALEPH2,PICH} \footnote{During the submission of this revised version of the paper, a
recent analysis \cite{PICH2} obtaining a value similar to ours has been published.}:
\beq
\overline{m}_s(1 GeV)\simeq (234^{+49_{exp}+32_{th}}_{-64_{exp}-37_{theo}}\pm 10.6_{fit}\pm
14.6_{J=0})~\mbox{MeV}~.
\eeq
However, our result is much more accurate due to the much better convergence of the QCD series in the
neutral vector current channel, but also to the more precise measurement of the sum rule in the neutral
channel where the contribution is dominated by the quite accurate measurement of the $\phi$ leptonic
width, and of the new accurate value of $R_{\tau,1}$ from recent ALEPH data.
Further improvements of  our result need much more accurate measurements of the $e^+e^-$ I=0 data or/and the
leptonic widths of the different $\phi$ families, but also improved control of the QCD perturbative series.
The agreement of the result with the one from
$\tau$-decay also indicates a good realization of the
$SU(2)$ symmetry relating the charged and  neutral currents (CVC). This result also agrees with some of
the results from the (pseudo)scalar sum rules \cite{SNB,SNR,YND}, \cite{SNP}--\cite{LEL} and from a global fit
\cite{DOSCH} of the light quark condensates from the hadron sector. The agreements of these different results
raise the questions of the reliability of the available estimates \cite{MALTA} of the parameters of the
$\omega$--$\rho$ mixing based on the SVZ sum rules
\cite{SVZ2}, and rule out a violent eventual reduction of about a factor 2 \cite{MALT} on the previous estimate in
Eq (\ref{d10}).
\section{On the $\omega$--$\rho$ mixing from the SVZ sum rules}
This quantity is controlled by the $SU(2)$-violating off-diagonal $\Pi_{08}$ correlator
defined in Eq. (7). In \cite{SVZ2}, SVZ have estimated this quantity using QCD spectral sum rules.
The sum rule reads:
\beq
\int dt~e^{-t\tau}\frac{1}{\pi}\Pi_{08}(t)\simeq \tau^{-1}\aga \delta_0+\delta_2\tau+\delta_4\tau^2
+\delta_6\tau^3\adr
\eeq
where:
\bea
\delta_0&=&\frac{\alpha}{16\pi^3},\nnb\\
\delta_2&=&\frac{3}{2\pi^2}(m^2_u-m^2_d)\nnb\\
\delta_4&=&\frac{md-m_u}{m_d+m_u}2m^2_\pi f^2_\pi\ga
1+\frac{md+m_u}{m_d-m_u}\frac{\gamma}{2+\gamma}\dr\nnb\\
\delta_6&=&-\frac{224}{81}\pi\rho\alpha_s\la\bar uu\ra^2\ga-\gamma+\frac{\alpha}{8\alpha_s}\dr~,
\eea
where $\alpha=1/137.036$ is the electromagnetic coupling and \cite{SNB}:
\beq
\gamma\equiv \la \bar dd\ra/\la\bar uu\ra\simeq 1-9\times 10^{-3}~.
\eeq
An analysis of the r.h.s of the sum rule shows that its $\tau$-stability is reached at small values of
$\tau\simeq 0.2~{\rm GeV}^{-2}$. This is a special feature compared with the case of the $\rho$-meson
sum rule, which stabilizes at the typical meson mass scale $\tau\simeq M_\rho^{-2}$, where the
lowest ground state contribution to the spectral integral dominates. 
This fact confirms SVZ previous arguments that the radial excitations $\omega'$--$\rho'$,... -mesons play an
important role in this analysis, and they should affect the lowest ground state contributions in a
significant way. The sensitivity of the analysis to the higher meson masses and to the finite
width effect of the $\rho$-meson should render
the estimate \cite{MALTA} (which comes from the cancellation of large numbers (see also \cite{NAS})) quite
inaccurate. These different features indicate that the estimate of the errors in the $\omega$--$\rho$-mixing
analysis
\cite{MALTA} should have been underestimated or/and the analysis needs a much better comprehensive study.
Effects of higher mass states should be more pronounced in the different finite energy sum rule (FESR) 
versions of the  Laplace transform sum rules proposed so far \cite{MALTA}, as well as in its
$\tau$-derivative (higher moments), which are needed in order to fix the complete set of the
$\omega$--$\rho$ mixing parameters.
\section{Summary and implications on the values of $\epsilon'/\epsilon$ and of $m_{u,d}$}
$\b$ We have estimated the strange-quark running mass using some new $\tau$-like $\phi$-meson sum rules, which 
are not affected by the eventual $SU(2)$ breaking due to the $\omega$--$\rho$-mixing. Our final result is
given in Eq. (\ref{final}), which is:
\beq
\overline{m}_s(1~{\rm GeV})\simeq (178\pm 33)~\mbox{MeV}~,\,\,\,\,\,\Longrightarrow
\,\,\,\,\,\overline{m}_s(2~{\rm GeV})\simeq (129\pm 24)~\mbox{MeV}~,
\eeq
while the positivity of the spectral function gives the upper bound in Eq. (\ref{bound}):
\beq
\overline{m}_s(1 ~{\rm GeV})\leq (200\pm 28)~\mbox{MeV}~,\,\,\,\,\,\Longrightarrow
\,\,\,\,\,\overline{m}_s(2~{\rm GeV})\leq (145\pm 20)~\mbox{MeV}~,
\eeq
$\b$ The agreement of this result with the ones given particularly in SN1 and updated in section 6,
and the different sum rules results from different channels \cite{SNB,SN1,ALEPH,PICH,YND}, \cite{SNP}--\cite{LEL},
rules out a huge effect due to the
$\omega$--$\rho$-mixing in the determination of $m_s$ from the
$\Delta_{10}$ sum rule, where this effect tends to reduce considerably the $m_s$ value \cite{MALT}.
Therefore, this result questions the validity of the different estimates  of the $\omega$--$\rho$-mixing
parameters based on the SVZ sum rule done in
\cite{MALTA}, which unlike other sum rules are very sensitive to the
high-energy behaviour of the spectral function, as we have seen previously.\\

\nin
$\b$ Our result in Eq. (\ref{final}) is in agreement within the errors, with the different results obtained
from various sum rule channels \cite{PDG,SNB,SNR} and lattice simulations in the quenched approximation
\cite{LATT,SNR}, though its central value is slightly higher than the sum rule plus chiral perturbation theory
`` grande average" \cite{SNR} of $\overline{m}_s(2~{\rm GeV})\simeq (119\pm 16)~\mbox{MeV}$. \\

\nin
$\b$ Due to the
difficulty for a precise determination of the non-perturbative four-quark matrix elements $Q_6\approx
B_6^{1/2}/m_s^2$ and $Q_8\approx B^{3/2}_8/m_s^2$ controlling the ratio of the CP-violating quantity
$\epsilon'/\epsilon$ \cite{BURAS}, which one expects to be more difficult than the extraction of $m_s$ from the
two-quark current correlator, we can combine the previous value of $m_s$ with the data on
$\epsilon'/\epsilon=(28\pm 4)\times 10^{-4}$ \cite{EPS}, in order to get an estimate of the combination
$Q_6-0.54Q_8$ within the standard model. Using the expression given in \cite{BURAS} and the value Im $\lambda_t
\simeq (1.34\pm 0.30)\times 10^{-4}$ from a standard analysis of the unitarity triangle, one
can easily deduce for the value of $m_s$ in Eq. (\ref{final}):
\beq
B_6^{1/2}-0.54 B^{3/2}_8 \simeq (2.8\pm 1.3)~,
\eeq
which is a useful phenomenological extraction of these matrix elements, and which can serve as
a guide in their non-perturbative determinations. This estimate may suggest a large violation of the vacuum
saturation estimate of $B_6^{1/2}\simeq B^{3/2}_8 \approx 1$ if negligible new physics affect the expression of
$\epsilon'/\epsilon$. However, such a violation may not be very surprising because analogous violations of the
vacuum saturation estimate of about a factor 2--3 has been already observed in the estimate of the four-quark
condensates within the sum rule approach of the meson and baryon channels and in the analysis of
the $e^+e^-$ and
$\tau$-decay data \cite{SNB,SNx,ALEPH2,OPAL}.\\

\nin
$\b$ Combining the result in Eq. (\ref{final}) with the one:
\beq
(\overline{m}_u+\overline{m}_d)(1~{\rm GeV})= (12\pm 2)~{\rm
MeV}~~~~\lrar ~~~~(\overline{m}_u+\overline{m}_d)(2~{\rm GeV})= (8.7\pm 1.5)~{\rm MeV},
\eeq
from the most recent pseudoscalar sum rule analysis \cite{RAFA,CNZ}, and from the direct
extraction of the light quark condensate \cite{DOSCH} \footnote{More complete discussions
can be found in \cite{SNR}.}, one obtains:
\beq
r_3\equiv\frac{2m_s}{m_u+m_d}= 30\pm 7~, 
\eeq
which is a non-trivial test of the chiral perturbation theory result of $24.4\pm 1.5$ \cite{LEUT}, that 
may be affected by some other ambiguities \cite{MANO}.\\

\nin
$\b$ Pursuing the update of the SN \cite{SN1} result, one can introduce the previous ratio into the Dashen
formula including NLO chiral corrections \cite{BIJ}:
\beq
r_2\equiv \frac{m_d-m_u}{m_d+m_u}=(0.52\pm 0.05)10^{-3}(r_3^2-1)\simeq 0.47\pm 0.16~,
\eeq
from which one obtains to order $a_s^3$:
\beq
(\overline{m}_d-\overline{m}_u)(1~{\rm GeV})= (5.6\pm 2.1)~{\rm MeV}~~~~\lrar
~~~~(\overline{m}_d-\overline{m}_u)(2~{\rm GeV})= (4.1\pm 1.5)~{\rm MeV}~,
\eeq
which one may compare with the lower bound $(\overline{m}_d-\overline{m}_u)(2~{\rm GeV})\geq 1.5~{\rm MeV}$
obtained in \cite{YND}. Solving the previous equations, one can deduce:
\bea
\overline{m}_d(1~{\rm GeV})= (8.8\pm 1.5)~{\rm MeV}~~~~\lrar
~~~~\overline{m}_d(2~{\rm GeV})= (6.4\pm 1.1)~{\rm MeV}~,\nnb\\
\overline{m}_u(1~{\rm GeV})= (3.2\pm 0.5)~{\rm MeV}~~~~\lrar
~~~~\overline{m}_u(2~{\rm GeV})= (2.3\pm 0.4)~{\rm MeV}~,
\eea
which we consider as an update of the previous sum rule results in \cite{SNP,SN1}.
\section*{Acknowledgements}
I thank, for its hospitality, the CERN Theory Division, where this work has been done, Kostja Chetyrkin for
communications, Graham Shore and Valya Zakharov for discussions. 
\vfill\eject

\end{document}